\documentclass[journal]{IEEEtran}
\ifCLASSINFOpdf
\else
\fi
\usepackage{graphicx}
\graphicspath{{figures/}} 
\usepackage{subcaption}

\usepackage{amsmath}

\usepackage{booktabs}
\usepackage{caption} 

\begin{document}
%
\title{SeisBind: Physics-Aware Tri-Modal Representation Binding for Seismic Data via Contrastive Learning}
%
%
%

\author{
\begin{minipage}[t]{0.45\textwidth}
\centering
Chaohua Liang\\
The University of Tokyo\\
Graduate School of Frontier Sciences\\
Kashiwa, Japan\\
\texttt{4532334173@edu.k.u-tokyo.ac.jp}
\end{minipage}
\hfill
\begin{minipage}[t]{0.45\textwidth}
\centering
Jun Matsushima\\
The University of Tokyo\\
Graduate School of Frontier Sciences\\
Kashiwa, Japan\\
\texttt{jun-matsushima@edu.k.u-tokyo.ac.jp}
\end{minipage}
}

%
%

\markboth{Preprint submitted to arXiv, January~2026}%
{Liang \MakeLowercase{\textit{et al.}}: SeisBind}
%



\maketitle

\begin{abstract}
\textbf{This letter proposes a physics-aware multi-modal contrastive learning framework designed to transform complex seismic wavefields into \textbf{human-readable physical representations}. Traditional data-driven inversion methods often focus on pixel-wise mapping, which lacks physical grounding and interpretability. To address this, we introduce a novel framework that jointly aligns seismic shot gathers, subsurface velocity models, and \textbf{explicit physical descriptors} (e.g., mean velocity and gradients) in a shared latent space. By introducing these descriptors as a third modality, our approach encourages the learned embeddings to capture intrinsic geological semantics rather than superficial signal correlations. Experiments on the OpenFWI dataset demonstrate that the proposed method not only achieves robust seismic-to-velocity retrieval but also effectively preserves meaningful physical semantics in the latent space. This representation-centric perspective enables a direct mapping from raw observations to interpretable physical attributes, providing a flexible foundation for expert-guided subsurface characterization.} 
\end{abstract}


%
\IEEEpeerreviewmaketitle

\section{Introduction}
%
%
%
%
\IEEEPARstart{S}{eismic} velocity models play a fundamental role in characterizing subsurface structures and serve as the basis for a wide range of geophysical exploration tasks, including seismic imaging, migration, and reservoir characterization. Accurate velocity information is essential for reliable interpretation of seismic data, as errors in velocity models can lead to distorted images and misleading geological conclusions. Consequently, estimating subsurface velocity models from seismic observations has long been a central problem in exploration geophysics.

Traditional approaches, primarily Full Waveform Inversion (FWI), are physics-driven but suffer from high computational costs and sensitivity to initial models \cite{virieuxOverviewFullwaveformInversion2009a}. While data-driven methods using Convolutional Neural Networks (CNNs) have emerged to accelerate estimation, most existing approaches rely on direct pixel-wise regression within a single modality. Consequently, they often lack physical interpretability and struggle to generalize to unseen geological scenarios \cite{zhangSeismicImpedanceInversion2022}. What's more, in conventional end-to-end deep learning frameworks, physical descriptors are difficult to be explicitly embedded into the learning process. As a result, the models tend to rely primarily on data-driven correlations and may overlook important physical information.

To address these challenges, we propose SeisBind, a physics-aware multi-modal contrastive learning framework. Instead of explicit inversion, we aim to learn a shared latent space where seismic data, velocity models, and statistical physical descriptors are semantically aligned. By introducing physical descriptors as a third modality, our approach acts as a semantic anchor, forcing the learned embeddings to capture intrinsic geological properties beyond superficial signal correlations. This representation-centric perspective enables robust cross-modal retrieval and provides a flexible foundation for expert-guided characterization.

Experimental results on the OpenFWI \cite{dengOPENFWILargescaleMultistructural} dataset  demonstrate that our method achieves precise seismic-to-velocity alignment and effectively preserves readable physical semantics. The main contributions of this work are:
\\1. We introduce a tri-modal contrastive learning framework that integrates physical descriptors with seismic and velocity data.
\\2. We demonstrate that explicit physical constraints significantly enhance the interpretability and quality of the learned representations.
\\3. We validate the proposed approach through cross-modal retrieval tasks, showing superior alignment performance.

\section{RELATED WORK}
Recent works utilized Convolutional Neural Networks (CNNs) to map seismic data directly to velocity models \cite{araya-poloDeeplearningTomography2018,kazeiMappingFullSeismic2021}. Other generative approaches, such as Generative Adversarial Networks (GANs), have also been employed for seismic impedance inversion with geophysical guidance \cite{zhangSeismicImpedanceInversion2022}. Besides, encoder-decoder architectures like InversionNet \cite{wuInversionNetEfficientAccurate2020} and other fully convolutional networks were developed to improve inversion accuracy and efficiency. More recently, researchers have explored incorporating velocity representation extensions to further enhance FWI performance \cite{muFullWaveformInversion2025}. While these methods reduce inference time, they often rely on pixel-wise regression, which may overlook global physical consistency.

Contrastive learning has revolutionized representation learning by aligning different views of data in a shared latent space, typically optimizing the InfoNCE loss \cite{oordRepresentationLearningContrastive2018}. The CLIP model \cite{radfordLearningTransferableVisual}  successfully demonstrated the power of aligning image and text modalities. This paradigm has been extended to various modalities, including audio \cite{guzhovAudioCLIPExtendingCLIP2021} and other sensory data, leading to unified embedding spaces like ImageBind \cite{girdharImageBindOneEmbedding2023} and UniBind \cite{lyuUniBindLLMAugmentedUnified2024}. These frameworks focus on learning transferable representations rather than solving a specific regression task, offering a new perspective for handling complex geophysical data.

In the geophysical domain, deep learning is increasingly combined with physics-based constraints. Kong et al. \cite{kongCombiningDeepLearning2022} demonstrated the benefits of combining deep learning with physics-based features for discrimination tasks. Recently, contrastive learning has been applied to subsurface geoscience to build vision-language models \cite{aseevContrastiveLearningBuilding2025} and foundation models like SeisCLIP \cite{siSeisCLIPSeismologyFoundation2023} for multi-purpose feature extraction. In parallel, generative approaches such as WaveDiffusion \cite{fengWaveDiffusion2026} have explored joint latent diffusion frameworks to synthesize seismic and velocity fields that approximately satisfy the governing wave equation.Different from these works, our approach explicitly introduces statistical physical descriptors as a third modality to constrain the learning process, this explicitly aligns the latent space, enabling the direct retrieval of interpretable physical parameters from raw seismic data or velocity models.

\section{METHODS}

\subsection{Problem Formulation}
In real seismic exploration, one of the main objectives is to calculate the subsurface velocity model using obtained seismic data. In this letter, we consider a paired seismic–velocity learning problem, where each training sample consists of three aligned components: (1) a seismic record, (2) a corresponding subsurface velocity model, and (3) a low-dimensional physics descriptor vector summarizing global physical properties of the model.
Instead of explicitly learning an inversion mapping from seismic data to velocity models, we aim to learn a shared embedding space in which seismic observations, velocity fields, and physics descriptors are semantically aligned. This formulation enables cross-modal retrieval and zero-shot generalization across acquisition and model distributions.

\subsection{Proposed Model}
\begin{figure*}[t]
    \centering
    \includegraphics[width=\textwidth]{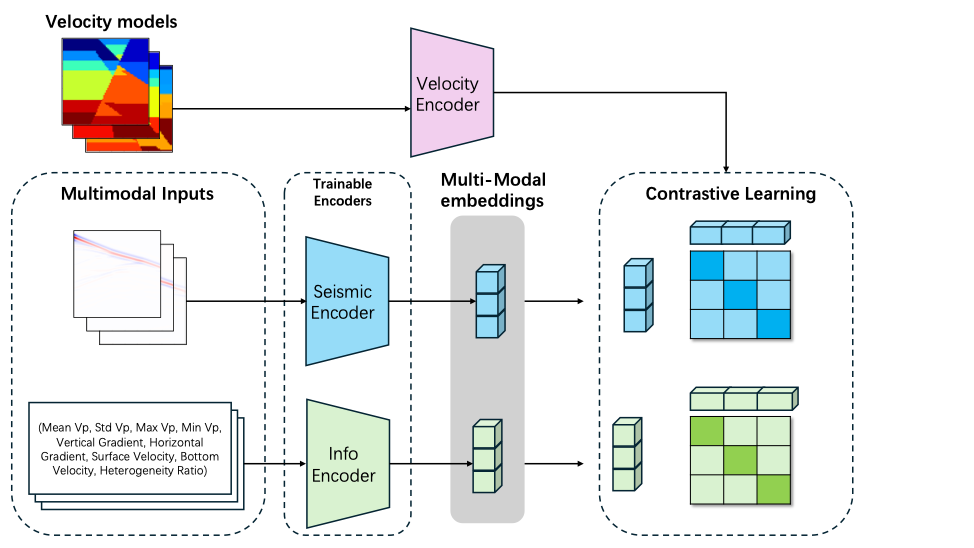}
    \caption{model architecture}
    \label{fig:model_arch}
\end{figure*}

The proposed model for multimodal alignment is illustrated in Fig. Inspired by contrastive multi-modal learning frameworks, we adopt a CLIP-style architecture composed of three modality-specific encoders: Velocity Encoder, Seismic Encoder and Info encoder, each encoder maps its input into a shared 256-dimensional embedding space and all embeddings are L2-normalized to lie on a unit hypersphere.

The input seismic tensor has shape (5,T,R), corresponding to five shot channels. The seismic encoder utilizes a symmetric four-block CNN architecture to extract latent features from multi-shot seismic gathers. Each block consists of a convolutional layer followed by batch normalization and ReLU activation. We employ standard square kernels ($7 \times 7$ to $3 \times 3$) with progressive downsampling to ensure a balanced receptive field across both temporal and spatial dimensions. The high-dimensional wavefield is finally aggregated via adaptive global average pooling and projected onto a 256-dimensional shared latent space, facilitating the alignment of raw observations with human-readable physical attributes.

The velocity encoder is implemented using a ResNet-18 backbone, which has been widely validated as a reliable and effective architecture for feature extraction \cite{heDeepResidualLearning2016}. Since the velocity input is single-channel, the first convolutional layer is modified accordingly. The final fully connected layer is replaced with a linear projection to the shared embedding dimension d.

To incorporate global physical priors, a lightweight multilayer perceptron (MLP) is employed to encode the physics vector $\mathbf{p}$. The network consists of three fully connected layers with ReLU activations. A batch normalization layer is applied at the input to mitigate scale discrepancies among different physics features.

The physics encoder acts as a semantic anchor, encouraging velocity embeddings to respect global physical consistency beyond pixel-level similarity.

\subsection{Contrastive Learning Objective}
We adopt an InfoNCE-based contrastive learning objective to align representations across seismic observations, velocity models, and physics constraints. Given embeddings $\mathbf{z}_i$ and $\mathbf{z}_j$, cosine similarity is used:

\begin{equation}
\operatorname{sim}(\mathbf{z}_i, \mathbf{z}_j) = \frac{\mathbf{z}_i^\top \mathbf{z}_j}{\|\mathbf{z}_i\| \|\mathbf{z}_j\|}
\label{eq:similarity}
\end{equation}

A symmetric CLIP-style loss \cite{oordRepresentationLearningContrastive2018,radfordLearningTransferableVisual} is applied to enforce bidirectional alignment:

\begin{equation}
\mathcal{L}_{i \rightarrow j} = - \frac{1}{N} \sum_{n=1}^{N} \log \frac{\exp\left( \operatorname{sim}\left( \mathbf{z}_i^{(n)}, \mathbf{z}_j^{(n)} \right) / \tau \right)}{\sum_{m=1}^{N} \exp\left( \operatorname{sim}\left( \mathbf{z}_i^{(n)}, \mathbf{z}_j^{(m)} \right) / \tau \right)}
\label{eq:contrastive_loss}
\end{equation}

where $\tau$ denotes a learnable temperature parameter. Instead of directly enforcing seismic–physics alignment, which suffers from a large representation gap and unstable gradients due to numerical sensitivity of physics residuals, we introduce velocity models as an intermediate physical representation. Accordingly, the final objective is defined as a weighted combination loss:

\begin{equation}
\mathcal{L} = \lambda_1 \mathcal{L}_{\text{Seismic--Velocity}} + \lambda_2 \mathcal{L}_{\text{Velocity--Physics}}
\label{eq:total_loss}
\end{equation}

This hierarchical contrastive strategy enables stable optimization while preserving physical consistency across modalities.

\section{EXPERIMENTS}

\subsection{Dataset and Processing}
In this letter, we use the opensource dataset OpenFWI, which provides large-scale synthetic seismic data and corresponding subsurface velocity models.. Specifically, we randomly selected 5000 samples from FlatFault and CurvedFault in OpenFWI dataset. Each sample in raw dataset includes seismic data with dimension of (5, 1000, 70), and correspond velocity model with dimension of (1, 70, 70). Physical descriptors with shape of (1 ,9) are deterministically extracted from velocity models through a set of statistical and structural operators, they encapsulate the statistical distribution (Mean/Std/Range), spatial variability (Gradients), and structural complexity (Heterogeneity Ratio) of the velocity model, providing explicit physical constraints to guide the multi-modal alignment process.

All seismic traces are normalized by their maximum absolute amplitude to stabilize training. Velocity models are linearly scaled using a fixed reference range to ensure numerical stability. The physics descriptors are normalized using batch normalization within the physics encoder.

\subsection{Experimental Setup}
The proposed framework consists of three modality-specific encoders: a seismic encoder, a velocity encoder, and a Info encoder. As illustrated above, the seismic encoder adopts a a symmetric CNN architecture, the velocity encoder is based on a ResNet-18 backbone adapted for single-channel input, while the Info encoder is implemented as a multilayer perceptron.

All three encoders project their respective inputs into a shared embedding space of dimension 256 and are trained jointly from scratch. A CLIP-style contrastive learning objective is employed to align embeddings across modalities. The model is optimized using the AdamW optimizer with a learning rate of 0.0001. What's more, we set $\lambda_1 = 1$ and empirically determine $\lambda_2 = 0.1$ to achieve the optimal trade-off between seismic--velocity alignment and physical consistency.

Training is performed for 10 epochs with a batch size of 128. All experiments are conducted on a single NVIDIA 2080Ti GPU. A learnable temperature parameter is introduced to scale the cosine similarity scores during contrastive training.

\subsection{Evaluation Metrics}
To evaluate the quality of the learned multimodal representations, we formulate cross-modal retrieval tasks between different modalities. Specifically, we consider (1) seismic-to-velocity retrieval and (2) velocity-to-physics retrieval. For each query sample, the goal is to retrieve its corresponding paired sample from a candidate pool containing all samples in the evaluation set.

Retrieval performance is quantified using Recall@$K$ ($R@K$), where $K \in \{1, 5, 10\}$. A retrieval is considered correct if the ground-truth paired sample appears within the top-$K$ retrieved candidates based on cosine similarity in the shared embedding space. This evaluation protocol directly measures the degree of alignment between different modalities.

\subsection{Results}
\begin{table}[htbp]
\centering
\caption{Retrieval Performance and Similarity Analysis.}
\label{tab:performance_results}
\begin{tabular}{@{}lcc@{}}
\toprule
\textbf{Metrics} & \textbf{Seismic $\rightarrow$ Velocity} & \textbf{Velocity $\rightarrow$ Physics} \\ \midrule
$R@1$ ($\uparrow$)            & 0.5780& 0.6810\\
$R@5$ ($\uparrow$)            & 0.8610& 0.9150\\
$R@10$ ($\uparrow$)           & 0.9350& 0.9610\\ \midrule
Mean Pos Sim ($\uparrow$)    & 0.6164& 0.6378\\
Mean Neg Sim ($\downarrow$)    & -0.0070& -0.0014\\ \bottomrule
\end{tabular}
\end{table}

Table I summarizes the quantitative retrieval results of the proposed method. The model achieves strong performance across all evaluated tasks, indicating an effective alignment between seismic data, velocity models, and physics descriptors in the shared embedding space.

In particular, seismic-to-velocity retrieval demonstrates high recall values, reflecting the model’s ability to associate complex seismic data with their corresponding subsurface velocity structures. Velocity-to-physics retrieval yields even higher recall scores, suggesting that the physics descriptors provide a compact yet informative representation of velocity models. Moreover, The model demonstrates exceptional modality alignment, as evidenced by a high Mean Positive Similarity alongside a near-zero Mean Negative Similarity. This substantial margin between positive and negative pairs indicates that the Tri-modal CLIP framework has learned a highly discriminative shared latent space, effectively suppressing inter-sample interference while capturing the intrinsic physical correlations between seismic responses and subsurface parameters.

\begin{table}[htbp]
\centering
\caption{Retrieval Performance Comparison: U-Net vs. SeisBind}
\label{tab:comparison}
\begin{tabular}{lccc}
\hline
\textbf{Metric} & \textbf{U-Net (Baseline)} & \textbf{SeisBind (Ours)} & \textbf{Improvement} \\ \hline
$R@1 \uparrow$  & 0.3270                    & \textbf{0.5780}          & +76.8\%             \\
$R@5 \uparrow$  & 0.5140                    & \textbf{0.8610}          & +67.5\%             \\
$R@10 \uparrow$ & 0.5840                    & \textbf{0.9350}          & +60.1\%             \\ \hline
\end{tabular}
\end{table}

For comparison, we tested the performance of the traditional U-Net model under the same test and compared it with SeisBind. Table II shows the comparative result for SeisBind and U-Net baseline. Specifically, SeisBind achieves $R@1$, $R@5$, and $R@10$ of 0.5780, 0.8610, and 0.9350, representing substantial improvements of 76.8\%, 67.5\%, and 60.1\% over the U-Net results (0.3270, 0.5140, and 0.5840). While traditional data-driven inversion methods often rely on pixel-wise mapping, SeisBind jointly aligns seismic data, physical descriptors and velocity models in a shared latent space, which enables the framework to capture intrinsic geological semantics. Besides, due to its unimodal and task-specific design, the U-Net baseline cannot be directly extended to multimodal retrieval tasks. For example, an inversion-based U-Net trained to map seismic shot gathers to velocity models cannot be directly applied to the reverse setting, i.e., inferring seismic observations from velocity models, without redefining the architecture and training objective, whereas SeisBind naturally supports cross-modal queries without architectural modification.

\begin{figure}[t]
  \centering
  \includegraphics[width=\linewidth]{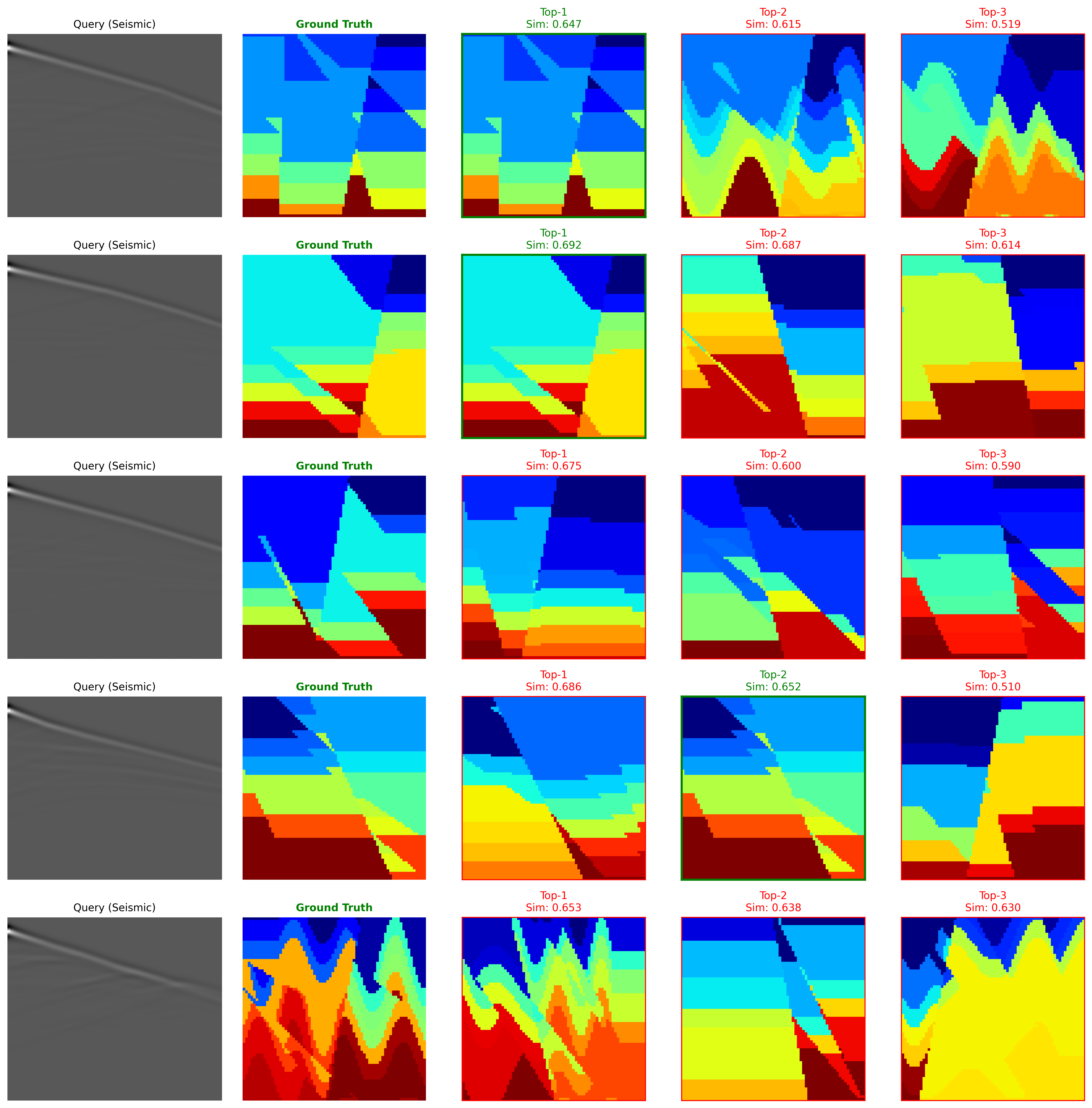}
  \caption{Qualitative retrieval results. For each query seismic shot gather (first column), the corresponding ground truth (second column) and the top-3 retrieved velocity models are displayed. Similarity scores are provided above each retrieved candidate, demonstrating the model's precision in structural matching.}
  \label{fig:topk_retrieval}
\end{figure}

To qualitatively evaluate the capability of Seisbind model, Figure 2 presents representative examples of the seismic-to-velocity retrieval results. As observed in the figure, the Top-1 retrieved velocity models exhibit high structural fidelity to the ground truth, accurately capturing key geological features such as layer interfaces and fault orientations. Even when the Top-1 match shows minor deviations, the Top-2 and Top-3 candidates still maintain consistent physical characteristics, confirming that the learned shared latent space effectively encodes the intrinsic correlations between seismic data and subsurface velocity models. This visual evidence aligns with the high recall values and the clear margins observed in the similarity matrices, further validating the robustness of the Tri-modal CLIP framework in associating diverse modalities.

\begin{table}[t]
\centering
\caption{Physics Parameter Inference from Seismic Embeddings via Cross-Modal Retrieval (k=5)}
\label{tab:physics_inference}
\begin{tabular}{lcc}
\toprule
\textbf{Physical Descriptor} & \textbf{MAE} & \textbf{Rel. Error (\%)} \\
\midrule
Surface Velocity   & 49.5 m/s    & 2.47  \\
Max Velocity       & 142.0 m/s   & 3.35  \\
Min Velocity       & 71.5 m/s    & 4.09  \\
Mean Velocity      & 125.9 m/s   & 4.10  \\
Bottom Velocity    & 185.6 m/s   & 4.54  \\
Heterogeneity Ratio & 0.022      & 8.14  \\
Std Velocity       & 85.5 m/s    & 10.35 \\
Vertical Gradient  & 6.39        & 13.91 \\
Horizontal Gradient & 15.9       & 31.14 \\
\midrule
\textbf{Average}   & \textbf{75.8} & \textbf{9.12} \\
\bottomrule
\end{tabular}
\end{table}

To validate that the learned embeddings preserve interpretable physical 
semantics, we perform physics descriptor inference via cross-modal retrieval. 
Given a seismic embedding, we retrieve its $k$ nearest velocity embeddings 
and average their associated physics vectors. Table~\ref{tab:physics_inference} 
shows the results using $k=5$ neighbors on the validation set.

Our method achieves an average relative error of \textbf{9.12\%}, with six 
parameters (surface, max, min, mean, and bottom velocities, plus heterogeneity) 
achieving errors below 5\%. Notably, surface velocity prediction reaches 
\textbf{2.47\%} error, demonstrating that physics-constrained embeddings 
effectively capture global statistical properties. While spatially-varying 
features like horizontal gradient show higher errors (31.14\%), this is 
expected as gradients are inherently local features that may not be fully 
captured by global embeddings. Overall, the performance is comparable to 
traditional physics-based inversion methods (typically 5--10\% error) while 
offering significantly faster inference.

To further analyze the learned representations, we visualize the similarity matrices between different modalities. Fig. 2 and 3 shows the cosine similarity heatmaps for seismic–velocity and velocity–physics pairs. Correctly matched pairs exhibit significantly higher similarity scores along the diagonal, while mismatched pairs remain close to zero, indicating good separation between positive and negative samples.

\begin{figure}[htbp] 
    \centering
    \includegraphics[width=0.9\columnwidth]{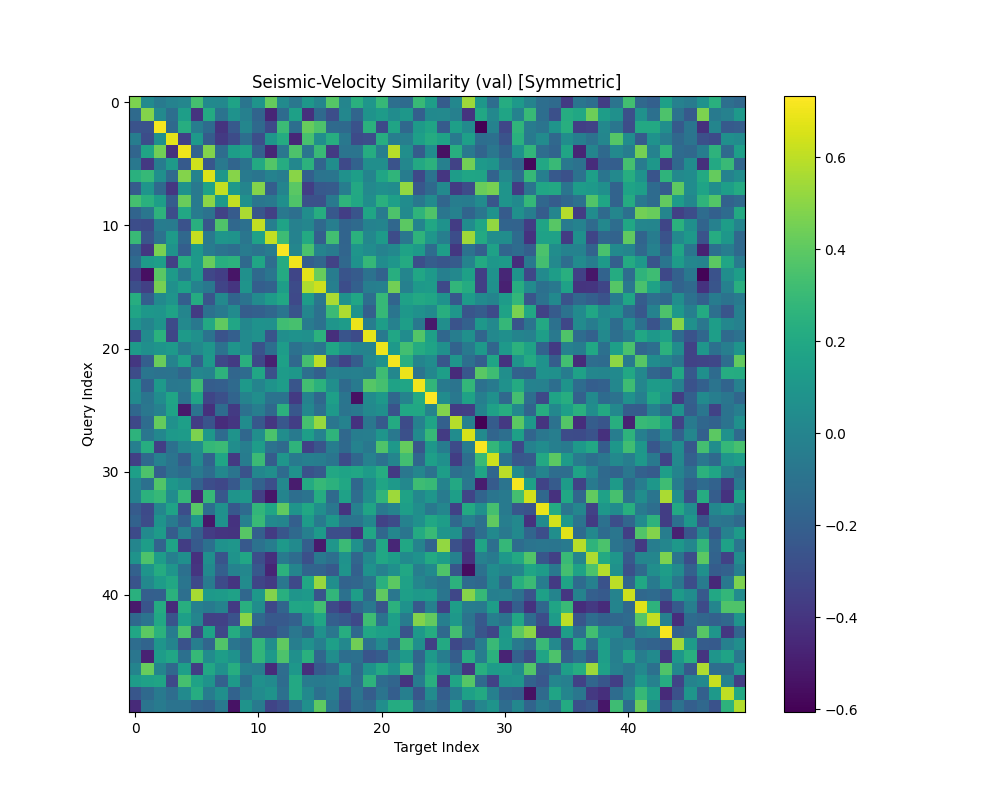}
    \caption{Similarity heatmap: Seismic-to-Velocity alignment.}
    \label{fig:heatmap_sv_vertical}
    
    \vspace{0.5cm} 
    
    \includegraphics[width=0.9\columnwidth]{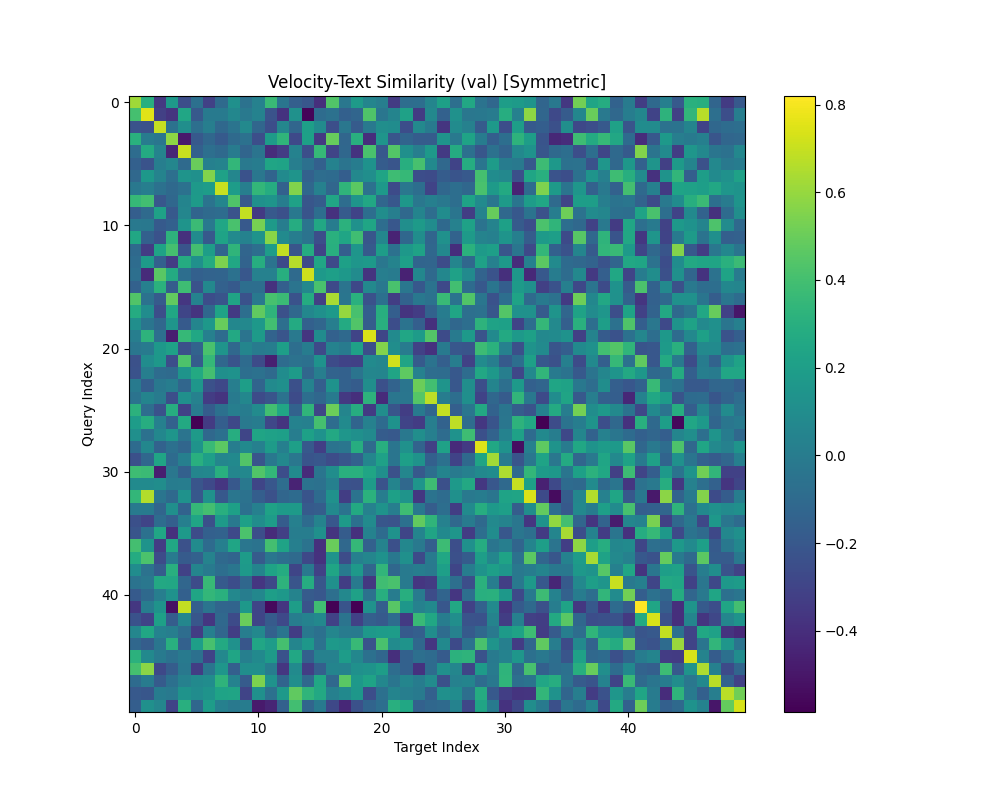}
    \caption{Similarity heatmap: Velocity-to-Physics alignment.}
    \label{fig:heatmap_vt_vertical}
\end{figure}

\section{Conclusion}
In this letter, we presented a novel physics-aware multi-modal contrastive learning framework for seismic representation learning. By integrating seismic shot gathers, velocity models, and explicit physical descriptors into a unified latent space, our approach provides new paradigms for seismic exploration. Experimental results on the OpenFWI dataset demonstrate that the proposed tri-modal alignment strategy not only achieves reliable seismic-to-velocity retrieval performance but also significantly enhances the physical interpretability of the learned embeddings. More importantly, the introduction of the physical modality acts as a semantic anchor, enabling the transformation of complex seismic data into human-readable attributes. This research provides a promising direction for seismic exploration. In future work, the SeisBind framework could be further extended to larger and more diverse datasets, while incorporating additional modalities and exploring a broader range of downstream tasks.


%


\section*{Acknowledgment}

The author gratefully acknowledges Prof. Jun Matsushima for his invaluable guidance and continuous support throughout this work.

\ifCLASSOPTIONcaptionsoff
  \newpage
\fi

\bibliographystyle{ieeetr}   
\bibliography{bibtex/bib/Clip}

\end{document}